\documentclass{amsart}
\usepackage{graphicx}
\usepackage{amscd}
\usepackage{amsmath}
\usepackage{amsfonts}
\usepackage{amssymb}
\newlength{\customskipamount}
\newlength{\customleftmargin}
\newbox\ipbox
\newcommand{\ip}[2]{\left\langle #1\mathrel{\mathchoice
{\setbox\ipbox=\hbox{$\displaystyle \left\langle\mathstrut #1#2\right\rangle$}
\vrule height\ht\ipbox width0.25pt depth\dp\ipbox}
{\setbox\ipbox=\hbox{$\textstyle \left\langle\mathstrut #1#2\right\rangle$}
\vrule height\ht\ipbox width0.25pt depth\dp\ipbox}
{\setbox\ipbox=\hbox{$\scriptstyle \left\langle\mathstrut #1#2\right\rangle$}
\vrule height\ht\ipbox width0.25pt depth\dp\ipbox}
{\setbox\ipbox=\hbox{$\scriptscriptstyle \left\langle\mathstrut #1#2\right\rangle$}
\vrule height\ht\ipbox width0.25pt depth\dp\ipbox}
} #2\right\rangle}

\begin{document}
\textbf{\textsc{Osterwalder--Schrader axioms---Wightman Axioms}}---The
mathematical axiom systems for quantum field theory (QFT) grew out of
Hilbert's sixth problem \cite{W}, that of stating the problems of quantum
theory in precise mathematical terms. There have been several competing
mathematical systems of axioms, and here we shall deal with those of A.S.
Wightman \cite{SW} and of K.~Osterwalder and R.~Schrader \cite{OS}, stated in
historical order. They are centered around group symmetry, relative to unitary
representations of Lie groups in Hilbert space. We also mention how the
Osterwalder--Schrader axioms have influenced the theory of unitary
representations of groups, making connection with \cite{JO}. Wightman's axioms
involve: (1)~a unitary representation $U$ of $G:=\mathrm{SL}\left(
2,\mathbb{C}\right)  \rtimes\mathbb{R}^{4}$ as a cover of the Poincar\'{e}
group of relativity, and a vacuum state vector $\psi_{0}^{{}}$ fixed by the
representation, (2)~quantum fields $\varphi_{1}^{{}}\left(  f\right)
,\dots,\varphi_{n}^{{}}\left(  f\right)  $, say, as operator-valued
distributions, $f$ running over a specified space of test functions, and the
operators $\varphi_{i}^{{}}\left(  f\right)  $ defined on a dense and
invariant domain $D$ in $\mathbf{H}$ (the Hilbert space of quantum states),
and $\psi_{0}^{{}}\in D$, (3)~a transformation law which states that $U\left(
g\right)  \varphi_{j}^{{}}\left(  f\right)  U\left(  g^{-1}\right)  $ is a
finite-dimensional representation $R$ of the group $G$ acting on the fields
$\varphi_{i}^{{}}\left(  f\right)  $, i.e., $\sum_{i}R_{ji}\left(
g^{-1}\right)  \varphi_{i}^{{}}\left(  g\left[  f\right]  \right)  $, $g$
acting on space-time and $g\left[  f\right]  \left(  x\right)  =f\left(
g^{-1}x\right)  $, $x\in\mathbb{R}^{4}$. (4)~The fields $\varphi_{j}^{{}%
}\left(  f\right)  $ are assumed to satisfy locality and one of the two
canonical commutation relations of $\left[  A,B\right]  _{\pm}=AB\pm BA$, for
fermions, resp., bosons; and (5)~it is assumed that there is scattering with
asymptotic completeness, in the sense $\mathbf{H}=\mathbf{H}^{\text{in}%
}=\mathbf{H}^{\text{out}}$.

The Wightman axioms were the basis for many of the spectacular developments in
QFT in the seventies, see, e.g., \cite{GJ1,GJ2}, and the Osterwalder--Schrader
axioms \cite{JO,OS} came in response to the dictates of path space measures.
The constructive approach involved some variant of the Feynman measure. But
the latter has mathematical divergences that can be resolved with an analytic
continuation so that the mathematically well-defined Wiener measure becomes
instead the basis for the analysis. Two analytical continuations were
suggested in this connection: in the mass-parameter, and in the
time-parameter, i.e., $t\mapsto\sqrt{-1}t$. With the latter, the Newtonian
quadratic form on space-time turns into the form of relativity, $x_{1}%
^{2}+x_{2}^{2}+x_{3}^{2}-t_{{}}^{2}$. We get a stochastic process
$\mathbf{X}_{t}$: symmetric, i.e., $\mathbf{X}_{t}\sim\mathbf{X}_{-t}$;
stationary, i.e., $\mathbf{X}_{t+s}\sim\mathbf{X}_{s}$; and
Osterwalder--Schrader positive, i.e., $\int_{\Omega}f_{1}\circ\mathbf{X}%
_{t_{1}}\,f_{2}\circ\mathbf{X}_{t_{2}}\,\cdots\,f_{n}\circ\mathbf{X}_{t_{n}%
}\,dP\geq0$, $f_{1},\dots,f_{n}$ test functions, $-\infty<t_{1}\leq t_{2}%
\leq\dots\leq t_{n}<\infty$, and $P$ denoting a path space measure.

Specifically: If $-t/2<t_{1}\leq t_{2}\leq\dots\leq t_{n}<t/2$, then
\begin{multline}%
\ip{\Omega}{A_{1}e^{-\left( t_{2}-t_{1}\right) \hat
{H}}A_{2}e^{-\left( t_{3}-t_{2}\right) \hat{H}}A_{3}\cdots A_{n}\Omega}%
\label{eqConsiderationsNew.1}\\
=\lim_{t\rightarrow\infty}\int\prod_{k=1}^{n}A_{k}\left(  q\left(
t_{k}\right)  \right)  \,d\mu_{t}\left(  q\left(  \,\cdot\,\right)  \right)  .
\end{multline}
By Minlos' theorem, there is a measure $\mu$ on $\mathcal{D}^{\prime}$ such
that
\begin{equation}
\lim_{t\rightarrow\infty}\int e^{iq\left(  f\right)  }\,d\mu_{t}\left(
q\right)  =\int e^{iq\left(  f\right)  }\,d\mu\left(  q\right)  =:S\left(
f\right)  \label{eqConsiderationsNew.2}%
\end{equation}
for all $f\in\mathcal{D}$. Since $\mu$ is a positive measure, we have
\[
\sum_{k}\sum_{l}\bar{c}_{k}c_{l}S\left(  f_{k}-\bar{f}_{l}\right)  \geq0
\]
for all $c_{1},\dots,c_{n}\in\mathbb{C}$, and all $f_{1},\dots,f_{n}%
\in\mathcal{D}$. When combining (\ref{eqConsiderationsNew.1}) and
(\ref{eqConsiderationsNew.2}), we note that this limit-measure $\mu$ then
accounts for the time-ordered $n$-point functions which occur on the left-hand
side in formula (\ref{eqConsiderationsNew.1}). This observation is further
used in the analysis of the stochastic process $\mathbf{X}_{t}$,
$\mathbf{X}_{t}\left(  q\right)  =q\left(  t\right)  $. But, more importantly,
it can be checked from the construction that we also have the following
reflection positivity: Let $\left(  \theta f\right)  \left(  s\right)
:=f\left(  -s\right)  $, $f\in\mathcal{D}$, $s\in\mathbb{R}$, and set
\[
\mathcal{D}_{+}=\left\{  f\in\mathcal{D}\mid f\text{ real valued, }f\left(
s\right)  =0\text{ for }s<0\right\}  \,.
\]
Then
\[
\sum_{k}\sum_{l}\bar{c}_{k}c_{l}S\left(  \theta\left(  f_{k}\right)
-f_{l}\right)  \geq0
\]
for all $c_{1},\dots,c_{n}\in\mathbb{C}$, and all $f_{1},\dots,f_{n}%
\in\mathcal{D}_{+}$, which is one version of Osterwalder--Schrader positivity. 

Since the Killing form of Lie theory may serve as a finite-dimensional metric,
the Osterwalder--Schrader idea \cite{OS} turned out also to have implications
for the theory of unitary representations of Lie groups. In \cite{JO}, the
authors associate to Riemannian symmetric spaces $G/K$ of tube domain type, a
duality between complementary series representations of $G$ on one side, and
highest weight representations of a $c$-dual $G^{c}$ on the other side. The
duality $G\leftrightarrow G^{c}$ involves analytic continuation, in a sense
which generalizes $t\mapsto\sqrt{-1}t$, and the reflection positivity of the
Osterwalder--Schrader axiom system. What results is a new Hilbert space where
the new representation of $G^{c}$ is ``physical'' in the sense that there is
positive energy and causality, the latter concept being defined from certain
cones in the Lie algebra of $G$. 

A unitary representation $\pi$ acting on a Hilbert space $\mathbf{H}(\pi)$ is
said to be \textit{reflection symmetric} if there is a unitary operator
$J:\mathbf{H}(\pi)\rightarrow\mathbf{H}(\pi)$ such that \begin{list}%
{}{\setlength{\leftmargin}{\customleftmargin}
\setlength{\itemsep}{0.5\customskipamount}
\setlength{\parsep}{0.5\customskipamount}
\setlength{\topsep}{\customskipamount}
\setlength{\parindent}{0pt}}
\item[\hss\llap{\rm R1)}] ${\displaystyle J^2 = \mbox{\rm id}}$.
\item[\hss\llap{\rm R2)}] ${\displaystyle J\pi(g) = \pi(\tau(g))J\, ,
\quad g\in G}$,
\end{list}
where $\tau\in\operatorname{Aut}\left(  G\right)  $, $\tau^{2}%
=\operatorname*{id}$, and $H:=\left\{  g\in G\mid\tau\left(  g\right)
=g\right\}  $.

A closed convex cone $C\subset\frak{q}$ is \textit{hyperbolic} if $C^{o}%
\not =\emptyset$ and if $\operatorname{ad}X$ is semisimple with real
eigenvalues for every $X\in C^{o}$.

Assume the following for $(G,\pi,\tau,J)$: \begin{list}{}{\setlength
{\leftmargin}{\customleftmargin}
\setlength{\itemsep}{0.5\customskipamount}
\setlength{\parsep}{0.5\customskipamount}
\setlength{\topsep}{\customskipamount}}
\item[\hss\llap{{\rm PR1)}}] $\pi
$ is reflection symmetric with reflection $J$.
\item[\hss\llap{{\rm PR2)}}] There is an $H$-invariant hyperbolic cone
$C\subset\frak{q}$ such that $S(C) = H\exp C$ is a closed semigroup and
$S(C)^o = H\exp C^o$ is diffeomorphic to $H\times C^o$.
\item[\hss\llap{{\rm PR3)}}] There is a subspace ${0}\not=
\mathbf{K}_0\subset\mathbf{H}(\pi
)$ invariant under $S(C)$ satisfying the positivity
condition
\[ \ip{v}{v}_J:= \ip{v}{J(v)} \ge0,\quad\forall v\in\mathbf{K}_0\, .\]
\end{list}

Assume that $(\pi,C,\mathbf{H},J)$ satisfies \textup{(PR1)--(PR3).} Then the
following hold: \begin{list}{}{\setlength{\leftmargin}{\customleftmargin}
\setlength{\itemsep}{0.5\customskipamount}
\setlength{\parsep}{0.5\customskipamount}
\setlength{\topsep}{\customskipamount}
\setlength{\parindent}{0pt}}
\item[\hss\llap{\rm1)}] $S(C)$ acts via
$s\mapsto\tilde{\pi}(s)$ by contractions on $\mathbf{K}$
($=$ the Hilbert space obtained
by completion of $\mathbf{K}_{0}$ in
the norm from (PR3)).
\item[\hss\llap{\rm2)}] Let $G^{c}$ be the simply
connected Lie group with Lie algebra $\frak{g}^{c}$. Then there exists a
unitary representation $\tilde{\pi}^{c}$ of $G^{c}$ such that
$d\tilde{\pi}^{c}(X)=d\tilde{\pi}(X)$ for $X\in\frak{h}$ and
$i\,d\tilde{\pi}^{c}(Y)=d\tilde{\pi}(iY)$ for $Y\in C$, where
$\frak{h}:=\left\{ X\in\frak{g}\mid\tau\left( X\right) =X\right\} $.
\item[\hss\llap{\rm3)}] The representation
$\tilde{\pi}^{c}$ is irreducible if and only if $\tilde{\pi}$ is irreducible.
\end{list}

\noindent Palle E.T. Jorgensen: \texttt{jorgen@math.uiowa.edu}

\noindent Gestur \'{O}lafsson: \texttt{olafsson@math.lsu.edu}
\end{document}